\documentclass[aps,prl,superscriptaddress,twocolumn, 10pt]{revtex4-1}
\usepackage{graphics}      
\usepackage{graphicx}      
\usepackage{url}           
\usepackage{bm}            
\usepackage{physics}
\usepackage{amsfonts}
\usepackage{float}
\usepackage{soul}
\usepackage{mathrsfs}
\usepackage{subcaption}
\usepackage{hyperref}
\usepackage{MnSymbol}
\usepackage{booktabs}
\usepackage{caption}
\usepackage{tikz,pgfplots}
\usetikzlibrary{backgrounds,fit,decorations.pathreplacing}
\usepackage{verbatim}
\usepackage{multirow}

\usepackage{siunitx} 

\usepackage[font=small,labelfont=rm]{caption}
\def \be {\begin{equation}}
\def \ee {\end{equation}}

\begin{document}
\title{Evidence of the entanglement constraint on wave-particle duality using the IBM Q quantum computer}
\author         {Nicolas Schwaller}
\email          {nicolas.schwaller@epfl.ch}
\affiliation{Institute of Physics, \'{E}cole Polytechnique F\'{e}d\'{e}rale de Lausanne (EPFL), Lausanne, CH-1015, Switzerland}
\author         {Marc-Andr\'e Dupertuis}
\email          {marc-andre.dupertuis@epfl.ch}
\affiliation{Institute of Physics, \'{E}cole Polytechnique F\'{e}d\'{e}rale de Lausanne (EPFL), Lausanne, CH-1015, Switzerland}
\author         {Cl\'ement Javerzac-Galy}
\email          {cjg@miraex.com}
\affiliation{Institute of Physics, \'{E}cole Polytechnique F\'{e}d\'{e}rale de Lausanne (EPFL), Lausanne, CH-1015, Switzerland}
\affiliation{Miraex, EPFL Innovation Park, B\^{a}timent L, Lausanne, CH-1015, Switzerland}
\date{\today}

\begin{abstract}
We experimentally verify the link existing between entanglement and the amount of wave-particle duality in a bipartite quantum system, with superconducting qubits in the IBM Q quantum computer. We consider both pure and mixed states, and study the influence of state purity on the observation of the complementarity ``triality'' relation of Jakob and Bergou. This work confirms the quantitative completion of local Bohr's complementarity principle by the nonlocal quantum entanglement typical of a truly bipartite quantum system.
\end{abstract}

\maketitle

\section{Introduction}
In 1924, physicist Louis De Broglie developed the theory of electron waves\,\cite{debroglie:tel-00006807}, coming up with the idea that particles behave like waves. This discovery is with no doubt one of the most stunning ideas in physics. Indeed, four years later, Niels Bohr formulated his principle of complementarity\,\cite{Bohr:1928:QPRb} dealing with this non intuitive property of Nature. It is possible to detect particle and wave characteristics of a single quantum object, but it never behaves fully like a wave and a particle at the same time. This idea was democratized by Richard Feynman in 1965, who underlined the strangeness of the so called \emph{wave-particle duality}: ``a phenomenon which is impossible, \emph{absolutely} impossible, to explain in any classical way, and which has in it the heart of quantum mechanics. In reality, it contains the \emph{only} mystery''\,\cite{feynman1965flp}. Experiments were in particular conducted with neutrons and photons (see e.g.\,\cite{RevModPhys.60.1067}), typically with double-slit setups, where a single quantum object has two possible paths, before to reach a screen where its position is measured. Knowing which path the object took indicates the object is a point-like particle, whereas observing an interference pattern on the screen, formed by the detected positions of the particles when the experiment is repeated, is the manifestation of the wave characteristic of the object, which apparently passes through both slits at once.

\

Wootters and Zurek initiated in 1979 a {\em quantitative} approach to wave-particle duality in a double slit experiment\,\cite{PhysRevD.19.473}, applicable to intermediate cases where the wave character would be incompletely revealed, and their surprising results prompted in 1988 a new and simpler mathematical description of Bohr's principle by Greenberger and Yasin\,\cite{GREENBERGER1988391}, namely
\be
V^2+P^2\leq 1  \, ,
\label{eq:greenberger-yasin_relation}
\ee
where $V, P \in [0, 1]$ are respectively the \emph{a priori} fringe \emph{visibility} and the which-way \emph{a priory distinguishability} (also called ``predictability''). $V$ is commonly associated to the waviness and $P$ to the particleness of a single quantum object. Equality holds for pure states, or ``coherent'' beams. Such type of inequality was later
investigated and extended by a few authors. Jaeger, Shimony, and Vaidman\,\cite{PhysRevA.51.54} proved a similar relationship in \emph{bipartite systems} (two-particle interferometer) by relating the visibility $V$ of one-particle interference fringes to the \emph{visibility of two-particle fringes} $V_{12} \in [0, 1]$ , i.e. $V^2+V_{12}^2\leq 1$. Englert\,\cite{PhysRevLett.77.2154} also obtained a similar result 
\be
V^2 + D^2\leq 1  \, ,
\label{eq:englert_relation}
\ee
where $D \in [0, 1]$ is this time the \emph{a posteriori distinguishability} (after detection, therefore also intrinsically bipartite). Equality holds when the which-path detector is similarly in a pure state. The analysis of the experiments for which relations of type (\ref{eq:greenberger-yasin_relation}) or (\ref{eq:englert_relation}) hold can thus be significantly different and easily induce misleading representations. Englert introduced a more prudent definition of the notion of (wave-particle) \emph{duality}, i.e. ``the observation of an interference pattern and the acquisition of which-way information are mutually exclusive'' and emphasized also that duality might not be enforced only by the use of position-momentum uncertainty relations as in the historical Bohr-Einstein debates (opening another debate, a recent account on this subtle aspect of complementarity can be found in Xiao et al.\cite{Xiaoeaav9547}). The second duality relation (\ref{eq:englert_relation}) was first experimentally tested by D\"{u}rr et al.\cite{PhysRevLett.81.5705} with an atom interferometer. For the first duality relation (\ref{eq:greenberger-yasin_relation}), earlier experiments in neutron interferometers implicitely tested it\,\cite{PhysRevA.36.4447,englert_bergou_2000}. For subsequent work Englert and Bergou\,\cite{PhysRevA.36.4447,englert_bergou_2000} include a short review of experimental and theoretical work on this topic (as of 2000), and put also on record a new erasure inequality quite similar to (\ref{eq:englert_relation}) but with different quantities outside our scope here.
Very recently Norrman et al.~\cite{Norrman2020} also derived interesting vector-light complementarity relations (\ref{eq:greenberger-yasin_relation}) and (\ref{eq:englert_relation}) in the case of double pinhole vectorial interference, the relevant visibility becomes then the Stokes visibility for polarization modulation.

\

Another important appreciation of wave-particle duality came with the realization of \emph{delayed-choice experiments}, first proposed by Wheeler\,\cite{RevModPhys.88.015005} as gedanken experiments. In such case the choice of the type of measurement is delayed at a later stage, which allows to challenge the idea that the measurement configuration could dictate \emph{a priori} the  waviness or particleness of the quantum system, as if the system would adapt to the choice of measurement. Delayed choice experiments were carried out for single particle by Jacques et al. in 2008\,\cite{PhysRevLett.100.220402} and for two particles by Ma et al. in 2009\,\cite{PhysRevA.79.042101}.
Since then a rich set of proposals and various experiments have been carried out: Wheeler's delayed-choice duality\,\cite{Marlow1978-MARMFO, Jacques966}, delayed-choice quantum erasure\,\cite{PhysRevA.25.2208, PhysRevLett.84.1, Mandel:91} and delayed-choice entanglement swapping\,\cite{doi:10.1080/09500340008244032} to mention a few. 

\

Despite such important and extensive progress it is quite obvious that duality relations envisaged so far, (\ref{eq:greenberger-yasin_relation}) and (\ref{eq:englert_relation}), are incomplete because they are inequalities, which can only bound duality. Indeed, if for example $V=1$ is measured then $P=0$, or vice versa, but if $V=0$ nothing can be deduced about the range of $P$. This highlights in a striking way the \emph{incompleteness} of duality relations (\ref{eq:greenberger-yasin_relation}) and (\ref{eq:englert_relation}). Things changed when people started to study quantitative complementarity occuring in composite systems (see\,\cite{jakob2003,PhysRevA.76.052107} and references therein). The simplest case is a bipartite composite quantum system composed of two qubits. Jakob and Bergou\,\cite{jakob2003, ref:bergou2010} have found in this case the single missing quantity which turned out to be the entanglement with the second qubit (defined by the concurrence $\mathscr{C}$), so that
\be
V^2 + P^2 + \mathscr{C}^2 = 1
\label{eq:jakob_bergou_relation}
\ee
is a ``triality'' relation and an equality which holds for any pure state of the two qubits. Furthermore this equation can be interpreted as a new complementarity relation between (wave-particle) duality of any of the two single qubit on one hand (the first two terms), and quantum entanglement with the second qubit on the other hand (the remaining term). So the amount of duality/complementarity in any of the two local subsystems determines the amount of bipartite non-local entanglement, the latter being also understood as a property which can exclude any of the two single-partite realities if $\mathscr{C}=1$. The beauty of this relationship resides in the fact that it relates the two most counter intuitive phenomenon of quantum physics, namely wave-particle duality and \emph{quantum entanglement}, in a single relation! Delayed-choice entanglement swapping experiments have now illustrated in a particularly bright manner such \emph{entanglement-separability duality} for bipartite (and multipartite) systems\,\cite{RevModPhys.88.015005}.

\

More recently Qian et al.\,\cite{Qian2018} derived a ``triality'' relation which looks totally similar to the Jakob-Bergou relation (\ref{eq:jakob_bergou_relation}). In fact the underlying mathematics is identical, thereby explaining the same identity, even though the physical content is quite different as two completely classical beams are considered, including the polarization degree of freedom for both beams. The analogy between the vector description of two classical polarized beams and a two-qubit quantum system is well known\,\cite{Spreeuw1998}, and obtained at the price of the introduction of a so-called ``position cebit'' together with the ``polarization cebit'' (standard Jones vector of one of the beams). The notion of ``classical entanglement'' that naturally ensues is still highly debated\,\cite{Karimi2015,ForbesEtAl2019}, and is sometimes referred to as ``entanglement of degrees of freedom'', ``single-particle entanglement'', or ``self-entanglement'', but it shows that quantum and classical physics do cross-fertilize again (see e.g.\cite{Spreeuw2001,Konrad2019} and references therein). While it lead some to argue that the quantum-classical boundary was shifting\,\cite{Qian2015}, we stress that in fact only the domain of application of Bell-like inequalities changes. 

\

Qian et al. also carried out a follow-up experiment in the quantum limit to verify a similar ``triality'' relation for single photons\,\cite{qian2019quantum}. In this regard it is necessary to point out that {\em single} photons can only test the {\em very same} classical structure of the field degrees of freedom: in a lossless linear optical system the transformation of single photon creation operators is the same as for the classical beam amplitudes (which also tells us that for more than one input photon other effects appear). So Ref.\,\cite{Qian2018} probes in fact {\em the same} triality relation as the classical experiment\,\cite{Qian2018}. This type of point of view was already exposed by Spreeuw\,\cite{Spreeuw1998}, we quote his conclusion ``The term classical entanglement seems justified even though a single particle is, strictly speaking, a quantum system. Single-photon entanglement is what remained when we took the low-intensity limit of a classical electromagnetic wave''.  

\

Another example of Jakob-Bergou relation involving entanglement of degrees of freedom is provided by a recent analysis of potential experiments with atom interferometers involving path and internal states of single atoms\,\cite{Miranda2020}.

\

In the present paper we offer the first experimental check of the {\em original} Jakob-Bergou ``triality'' relation (\ref{eq:jakob_bergou_relation}) for a {\em genuine bipartite} quantum system of two qubits, namely the superconducting qubits of the IBM Q quantum computer~\cite{ibm}, harnessed by the current fascinating progress in widely accessible quantum technologies. 

\section{Quantum waviness, particleness and entanglement}

Consider a general pure state of two qubits,
\be
\ket{\psi}=\alpha\ket{00}+\beta\ket{01}+\gamma\ket{10}+\delta\ket{11},
\label{eq:general_state}
\ee
with $\alpha, \beta, \gamma, \delta \in \mathbb{C}$ satisfying the normalization
\be
|\alpha|^2+|\beta|^2+|\gamma|^2+|\delta|^2=1.
\label{eq:pure_norm}
\ee
The state (\ref{eq:general_state}) can be characterized by its density matrix,
\be
\rho=\begin{pmatrix}
\alpha\alpha^* & \alpha\beta^* & \alpha\gamma^* & \alpha\delta^*\\
\beta\alpha^* & \beta\beta^* & \beta\gamma^* & \beta\delta^*\\ 
\gamma\alpha^* & \gamma\beta^* & \gamma\gamma^* & \gamma\delta^*\\ 
\delta\alpha^* & \delta\beta^* & \delta\gamma^* & \delta\delta^*
\end{pmatrix}.
\label{eq:rhoAB}
\ee

By convention, the first and second qubits will be respectively called qubit $A$ and qubit $B$. The corresponding reduced density matrices of subsystems $A$ and $B$ are
\be
\rho_{A}=Tr_B(\rho)=\begin{pmatrix}
\alpha\alpha^* + \beta\beta^* & \alpha\gamma^*+\beta\delta^*\\ 
\gamma\alpha^*+\delta\beta^* & \gamma\gamma^*+\delta\delta^*
\end{pmatrix}
\label{eq:rhoA}
\ee
and
\be
\rho_{B}=Tr_A(\rho)=\begin{pmatrix}
\alpha\alpha^* + \gamma\gamma^* & \alpha\beta^*+\gamma\delta^*\\ 
\beta\alpha^*+\delta\gamma^* & \beta\beta^*+\delta\delta^*
\end{pmatrix}.
\label{eq:rhoB}
\ee
Three central quantities\,\cite{jakob2003,ref:bergou2010} can then be derived.

\

First, the \emph{concurrence}, defined in the bipartite pure case by
\be
\mathscr{C}(\psi)=2|\alpha\delta-\beta\gamma|.
\label{eq:def_C}
\ee
The concurrence indicates the amount of entanglement between two quantum systems\,\cite{PhysRevLett.78.5022,wooters:eofandconcurrence} as it is a monotone of the \emph{entanglement of formation}, $E_f$, which is a measure of entanglement based on the separability criterion: $E_f=0$ if and only if the density matrix can be written as a mixture of product states. Both $\mathscr{C}$ and $E_f$ take the value one for maximally entangled states.

\

Second, the \emph{coherence} $\mathscr{V}_k$ between the two orthogonal states $\ket{0}$ and $\ket{1}$ of the qubit $k$, which is therefore a quantity related to a single qubit. It is directly proportional to the norm of the off-diagonal elements of its density matrix, and reads
\be
\mathscr{V}_k=2 |\rho_{k_{12}}|,\ \ \ k=A,B.
\label{eq:def_V}
\ee
Note that the counterpart of coherence in an interference experiment is the visibility.

\

Third, the \emph{predictability} $\mathscr{P}_k$, which quantifies the knowledge of ``which proportion'' of the system $k$ is in the state $\ket{0}$ or $\ket{1}$. It is defined by
\be
\mathscr{P}_k= |\rho_{k_{22}}-\rho_{k_{11}}|,\ \ \ k=A,B.
\label{eq:def_P}
\ee
The predictability is analogous to the which-path information in an interference experiment.

\

By replacing the definitions (\ref{eq:rhoAB}\,-\,\ref{eq:rhoB}) in Eqs.\,(\ref{eq:def_C}\,-\,\ref{eq:def_P}) it is easy to show that
\be
\mathscr{V}_k^2+\mathscr{P}_k^2+\mathscr{C}^2=(|\alpha|^2+|\beta|^2+|\gamma|^2+|\delta|^2)^2.
\label{eq:proof_VDC<1}
\ee
One notices that the right-hand side of (\ref{eq:proof_VDC<1}) is nothing else than the norm of the state (\ref{eq:general_state}) raised to the power 4. Thus, one can conclude that for a pure state\,\cite{jakob2003,ref:bergou2010},
\be
\mathscr{V}_k^2+\mathscr{P}_k^2+\mathscr{C}^2=1.
\label{eq:VPC=1}
\ee
Note that (\ref{eq:VPC=1}) remarkably claims that for a pure state of two qubits, the amount of entanglement strictly pilots the amount of duality of any qubit of the pair, namely $\mathscr{V}_k^2+\mathscr{P}_k^2$, $k=A,B$, which has the same value for both qubits. Conversely Eq.\,(\ref{eq:VPC=1}) also nicely reflect the well-known fact that local unitary transformations on any of the qubits cannot change the amount of mutual entanglement.

\section{Experiment on IBM Q}

\begin{figure}[H]
  \centerline{
    \begin{tikzpicture}[thick]
    \tikzstyle{operator} = [draw,fill=white,minimum size=1.5em] 
    \tikzstyle{phase} = [fill,shape=circle,minimum size=5pt,inner sep=0pt]
    \tikzstyle{surround} = [fill=blue!10,thick,draw=black,rounded corners=2mm]
    %
    \node at (-0.1,0) (q1) {$\ket{0}$};
    \node at (-0.1,-1) (q2) {$\ket{0}$};
    %
    \node[operator] (op11) at (1.2,0) {$\rm{R_y}(\alpha)$} edge [-] (q1);
    %
    \node[phase] (phase11) at (2.3,0) {} edge [-] (op11);
    \node[operator] (cu3) at (2.3,-1) {$\rm{R_y}(\theta)$} edge [-] (q2);
    \draw[-] (phase11) -- (cu3);
    %
    \draw[black,thick] (3.3,0.25) -- (5.3,0.25) -- (5.3,-1.25) -- (3.3,-1.25) -- (3.3,0.25) {};
    \node at (4.35,-0.3) {Linear};
    \node at (4.3,-0.7) {tomography};
    %
    \node (end1) at (3.4,0) {} edge [-] (op11);
    \node (end2) at (3.4,-1) {} edge [-] (cu3);
    \node (be1) at (5.2,0) {};
    \node (be2) at (5.2,-1) {};
    \node (ee1) at (5.75,0) {} edge [-] (be1);
    \node (ee2) at (5.75,-1) {} edge [-] (be2);
    \tikzset{meter/.append style={draw, inner sep=5, rectangle, font=\vphantom{A}, minimum width=20, line width=.8,path picture={\draw[black] ([shift={(.1,.15)}]path picture bounding box.south west) to[bend left=50] ([shift={(-.1,.15)}]path picture bounding box.south east);\draw[black,-latex] ([shift={(0,.1)}]path picture bounding box.south) -- ([shift={(.2,-.1)}]path picture bounding box.north);}}}
    \node[meter] (meter) at (6,0) {};
    \node[meter] (meter) at (6,-1) {};
     \begin{pgfonlayer}{background}
        \node at (1.7,0.7) {State preparation};
        \draw[black,thick,dashed] (0.35,0.5) -- (3.1,0.5) -- (3.1,-1.5) -- (0.35,-1.5) -- (0.35,0.5);
        \node[dashed] (background) [fit = (op11) (cu3)] {};
    \end{pgfonlayer}
    \end{tikzpicture}
  }
  \caption{Quantum circuit composed of two gates to prepare the initial state as a function of two parameters, followed by linear tomography circuits, and related measurements.}
  \label{fig:quantum_circuit}
\end{figure}
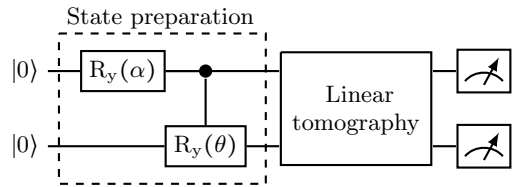

We create a tunable state on the Bloch sphere with the simple circuit shown in Fig.\,\ref{fig:quantum_circuit} and use linear tomography to obtain $\mathscr{V}_k$, $\mathscr{P}_k$ and $\mathscr{C}$, and check the Jakob-Bergou relation (\ref{eq:VPC=1}) [or the related inequality (\ref{eq:VPC_leq_1})].
\noindent In terms of the $\alpha$ and $\theta$ parameters, the preparation stage creates (see Appendix A)
\be
\ket{\psi}=\cos\frac{\alpha}{2}\ket{00}+\cos\frac{\theta}{2}\sin\frac{\alpha}{2}\ket{10}+\sin\frac{\theta}{2}\sin\frac{\alpha}{2}\ket{11}.
\label{eq:state_psi}
\ee
Then we perform a two-qubit state tomography, allowing to retrieve the density matrix of the quantum state, an intermediate step from which we compute $\mathscr{V}_k$, $\mathscr{P}_k$ and $\mathscr{C}$. The tomography procedure is the linear method proposed in \cite{PhysRevA.64.052312}, using the set of four Stokes measurements \{$\hat{\mu}_0=\ketbra{0}$, $\hat{\mu}_1=\ketbra{1}$, $\hat{\mu}_2=\ketbra{\bigplus}$, $\hat{\mu}_3=\ketbra{\circlearrowleft}$\} where $\ket{\bigplus}=\frac{1}{\sqrt{2}}(\ket{0}+\ket{1})$, $\ket{\circlearrowleft}=\frac{1}{\sqrt{2}}(\ket{0}+i\ket{1})$. Following linear tomography $\mathscr{V}_k$, $\mathscr{P}_k$ and $\mathscr{C}$ are retrieved via their direct link to the measured density matrix [Eqs.\,(\ref{eq:def_V}), (\ref{eq:def_P}) and (\ref{eq:expression_C})]. Although such a method is less direct than operational interferometric measurements of $\mathscr{V}_k$ and $\mathscr{P}_k$, it provides more accurate results since less quantum gates are required. The quantum nondemolition circuit proposed by \cite{PhysRevLett.98.250501} would also be necessary for an additional direct measurement of $\mathscr{C}$ (which is possible only for two-qubit states with real coefficient in the computational basis, as noted by \cite{PhysRevLett.98.250501}). This final measurement would also require more gates, as well as two additional ancillary qubits.

For a genuine test it is important to check the potential falsifiability of the Jakob-Bergou relation by our measurement procedure. One may object that Eq.\,(\ref{eq:VPC=1}) [or the related inequality (\ref{eq:VPC_leq_1})] mathematically follows from the density matrix properties, hence our procedure based on tomography cannot be a serious test. This would indeed be true if one would use {\em maximum likelyhood} tomography\,\cite{PhysRevA.64.052312} which automatically constructs perfect density matrices from a mathematical perspective, for which the Jakob-Bergou relation is always satisfied. However in our case we perform {\em linear} tomography which consists of the strict minimum of sixteen measurements necessary to unambiguously deduce the matrix elements of a $4\times 4$ hermitian matrix with normalized trace, expected to be the density matrix for two qubits, but there is no safety net that would ensure automatic non-negativity of this matrix (occurrence of non-positive matrices is well-known). Therefore non-physical results could also be produced for any physical observable subsequently computed with such matrices. The most likely reason is of course noise, but non-validity of quantum mechanics could also manifest as systematic violations, so in our case the test is genuine even if it is indirect. Most importantly it is as strained as possible due to the limited number of gates used. Furthermore we shall be able to check that among the results produced for $\mathscr{V}_k$, $\mathscr{P}_k$ and $\mathscr{C}$ there are rare violations of the Jakob-Bergou relation. Finally we shall see that these violations disappear with longer statistical averaging, so they are clearly attributable to statistical noise, as expected.

\begin{figure}[h]
\centering
  \includegraphics[width=0.42\textwidth]{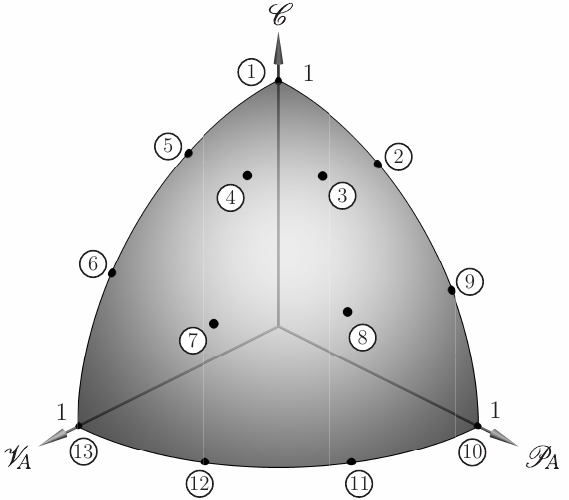}
  \caption{Analytical $\mathscr{V}_A$, $\mathscr{P}_A$ and $\mathscr{C}$ for the 13 pure states described by the angles listed in Tab.\,\ref{tab:values_alpha_theta} of Appendix A. All points lie on a sphere of unit radius.}
  \label{fig:VDCsim}
\end{figure}

To illustrate the equality (\ref{eq:VPC=1}) we display on Fig.\,\ref{fig:VDCsim} the values of $\mathscr{V}_A$, $\mathscr{P}_A$ and $\mathscr{C}$ which correspond to pure state of two qubits of the form (\ref{eq:state_psi}), and it covers the unit sphere belonging to the first octant. We also show the position of the points corresponding to the 13 states chosen in Appendix A, and which will be subsequently measured on IBMQ.

With the aim of performing the experiment on the real quantum computer, a \emph{noisy intermediate-scale quantum} (NISQ) computer\,\cite{Preskill2018quantumcomputingin}, formulas need to be extended to mixed states.
For a mixed state with density matrix
\be
\rho=\sum_{j}p_{j}\ket{\phi_{j}}\bra{\phi_{j}},
\label{eq:mixed_rho}
\ee
where $\ket{\phi_{j}}$ are pure states composing the complete state with probability $p_j$, it is possible to compute the concurrence\,\cite{PhysRevLett.80.2245} by defining the \emph{spin flip} matrix
\be
\Sigma=\sigma_y\otimes\sigma_y
\label{eq:matrix_Sigma}
\ee
and the matrix
\be
R(\rho)=\rho\Sigma\rho^*\Sigma.
\label{eq:matrix_R}
\ee
The concurrence is given by
\be
\mathscr{C}=\max(0,\sqrt{r_1}-\sqrt{r_2}-\sqrt{r_3}-\sqrt{r_4}).
\label{eq:expression_C}
\ee
where $r_1 \geq r_2 \geq r_3 \geq r_4$ are the eigenvalues of $R(\rho)$. Using expressions (\ref{eq:matrix_Sigma}) to (\ref{eq:expression_C}) allows to compute the concurrence of the pair of qubits from the linear tomography step.

\

The coherence of the qubit $k$ in the mixed bipartite case\,\cite{Tessier2005} is given by
\be
\mathscr{V}(\rho_k)=2\left|Tr(\rho_k\sigma_+^{(k)})\right|,
\label{eq:mixed_coherence_Tessier}
\ee
where $\sigma_+^{(k)}=\begin{pmatrix} 0 & 1\\
0 & 0 
\end{pmatrix}$ is the raising operator acting on qubit $k$.
\\
It can be written as
\be
\mathscr{V}_k=\sum_{i\neq j}\left|\rho_{k_{ij}}\right|,
\label{eq:mixed_coherence}
\ee
which, for a pure state, is equivalent to (\ref{eq:def_V}) thanks to the hermiticity of the density matrix.
Similarly, the predictability (of the state) of a qubit\,\cite{Tessier2005} is given by (\ref{eq:def_P}) in the case of a two-qubit mixed state.

\subsection{Experimental results}

\begin{figure}[h]
\multirow{2}{*}{}{
         \begin{subfigure}[h]{0.5\textwidth}
             \begin{center}
             \includegraphics[width=0.88\textwidth]{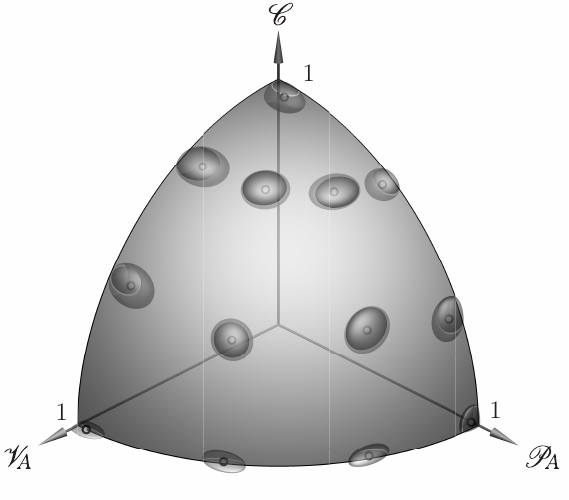}
             \caption{\centering}
             \label{fig:mitigated_1000_VPC}
             \end{center}
         \end{subfigure}\\
         \begin{subfigure}[h]{0.5\textwidth}
             \begin{center}
             \includegraphics[width=1\textwidth]{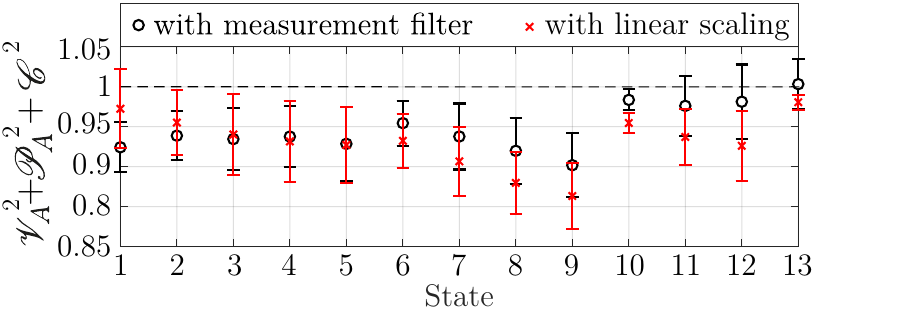}
             \caption{\centering}
             \label{fig:mitigated_1000_errorbar}
             \end{center}
         \end{subfigure}\\
         }
\caption{(a) Evaluations of $\mathscr{V}_A$, $\mathscr{P}_A$ and $\mathscr{C}$ for the states introduced in Fig.\,\ref{fig:VDCsim}, corrected with \emph{qiskit ignis} measurement filter. For visibility ellipsoids represent $3\sigma$ deviation from the mean. (b) Check of triality relation (\ref{eq:VPC_leq_1}) from the same data, error bars correspond this time to standard deviation. Results with measurement filter and linear scaling of $\mathscr{C}$ can be compared.}
\label{fig:norm_qx2}
\end{figure}

\begin{figure}[H]
\centering
  \centering
  \includegraphics[width=1.13\linewidth]{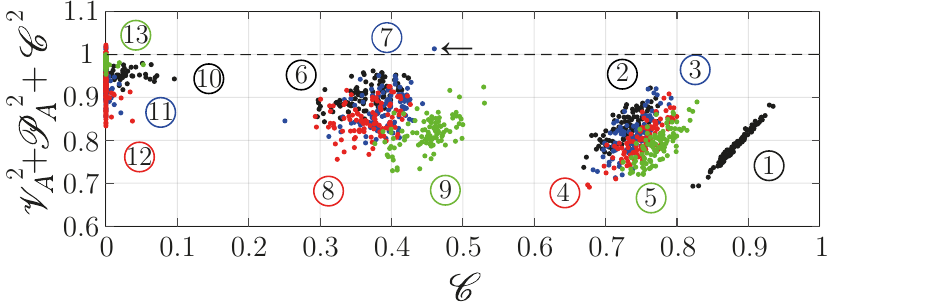}
  \caption{Raw values\,\cite{date_meas_1000} of $\mathscr{V}_A^2 + \mathscr{P}_A^2 + \mathscr{C}^2$ appearing in the triality relation (\ref{eq:VPC_leq_1}) for the 13 states introduced in Fig.\,\ref{fig:VDCsim}. The arrow shows a result violating the Jakob-Bergou inequality.}
  \label{fig:raw_data_1000}
\end{figure}

Given the possibility to compute the quantities for a mixed state, we can now perform the experiment with the real qubits. For this, the backend ibmq\_rome is used\,\cite{date_meas_1000}. We perform 1000 shots for each of the 16 linear tomography circuits used to compute a density matrix, and to evaluate subsequently $\mathscr{V}_k$, $\mathscr{P}_k$ and $\mathscr{C}$. Then each experiment is repeated 100 times to be able to evaluate the distribution of the results. Figure\,\ref{fig:mitigated_1000_VPC} shows the experimental results corresponding to Fig.\,\ref{fig:VDCsim}. 
In each case $3\sigma$-ellipsoids centered around mean values are represented and give an idea of the fitted normal distribution of $\mathscr{V}_A$, $\mathscr{P}_A$, and $\mathscr{C}$. 
In Figure\,\ref{fig:mitigated_1000_VPC}, we have used the error mitigation tool provided by \emph{qiskit ignis}\cite{Qiskit} for the two corresponding qubits. In this process, a \emph{measurement filter} is computed from the outcomes of the measurements of the four computational basis states. Following the toolbox, the filter is applied to the raw measurements outcomes of each of our circuits, before the density matrices are evaluated. This is supposed to compensate for the noise and decoherence taking place in the ibmq\_rome backend, and we see that indeed the results agree quite well with Fig.\,\ref{fig:VDCsim}. A closer look at the distance from Jakob-Bergou equality is provided by the complementary Figure~\ref{fig:mitigated_1000_errorbar} where $\mathscr{V}_A^2 + \mathscr{P}_A^2 + \mathscr{C}^2$ is reported for each state. We see that error mitigation is roughly equivalent to upscaling the $\mathscr{C}$-axis of the raw results by $\sim 1/0.899$, whilst leaving $\mathscr{V}_A$, $\mathscr{P}_A$ unchanged. On the equator, where entanglement between the two qubits is vanishing (not the most interesting part), the equality is best satisfied. For all other states with non-vanishing concurrence, we see that $\mathscr{V}_A^2 + \mathscr{P}_A^2 + \mathscr{C}^2$ clearly falls slightly below unity (this is even more true without error mitigation or upscaling, c.f. Fig.\,\ref{fig:raw_data_1000}). This is not surprising since the mixedness of the state (due to unavoidable experimental decoherence and noise) implies~\cite{jakob2003,ref:bergou2010}
\be
\mathscr{V}_k^2 + \mathscr{P}_k^2 + \mathscr{C}^2\leq 1 \quad (k=A,B).
\label{eq:VPC_leq_1}
\ee
In the next section we shall further prove that the limited purity of the generated state does exclusively explain the maximum observed concurrence level (and justifies the scaling factor). 

Fig.\,\ref{fig:raw_data_1000} offers a closer look at the 100 measured raw values corresponding to Fig.\,\ref{fig:mitigated_1000_VPC}, as a function of $\mathscr{C}$. Two features are noteworthy. First for the states with highest concurrence (states 1-4) we see elongated clouds which show that the higher the concurrence the better the equality in (\ref{eq:VPC_leq_1}). It is a manifestation of the fact that purity limits concurrence, as shown in the next section. The second feature worth noting is the existence of points clearly violating the Jakob-Bergou inequality (only one with $\mathscr{C}>0$, indicated by the arrow). In such case we checked that the intermediate density matrix is also non-positive (necessary but non-sufficient prerequisite). The frequency of such violations does strongly diminish with the number of shots used before the intermediate density matrix evaluation (and they are already rare for 1000 shots). This disappearance is in accordance with the assumption that they are due to noise, and prove that the Jakob-Bergou relation is valid, just as quantum mechanics. However their mere existence confirms the potential falsifiability of the Jakob-Bergou relation in our experiment. 

\subsection{Purity limits concurrence}

In order to improve and understand better the scaling observed between the experimental concurrence $\mathscr{C_{\exp}}$ and its theoretical counterpart $\mathscr{C}_{\rm{th}}$ we carry out a second experiment. Now a thousand states with random values of $\alpha$ and $\theta$ are generated and measured using the ibmq\_rome backend\,\cite{date_meas_8192}, this time with a larger number of 8192 shots per circuit. For each of these states the raw values of $\mathscr{V}_k$, $\mathscr{P}_k$ and $\mathscr{C}$ are displayed in Fig.\,\ref{fig:raw_data_random} in Appendix B (for qubits $A$ and $B$), and no violation of the Jakob-Bergou inequality is observed anymore. We have also measured a negative correlation (Pearson correlation coefficient of $-0.418$) between the state purity $\Tr(\rho^2)$ and the distance $\mathscr{C}_{\rm{th}}-\mathscr{C}_{\exp}$. This fact is not surprising, as entanglement is a fragile resource\,\cite{RevModPhys.81.865} to the environment, and concurrence is known to be limited by the state purity\,\cite{Ziman2005}.

\begin{figure}[h]
  \centering
  \includegraphics[width=1.0\linewidth]{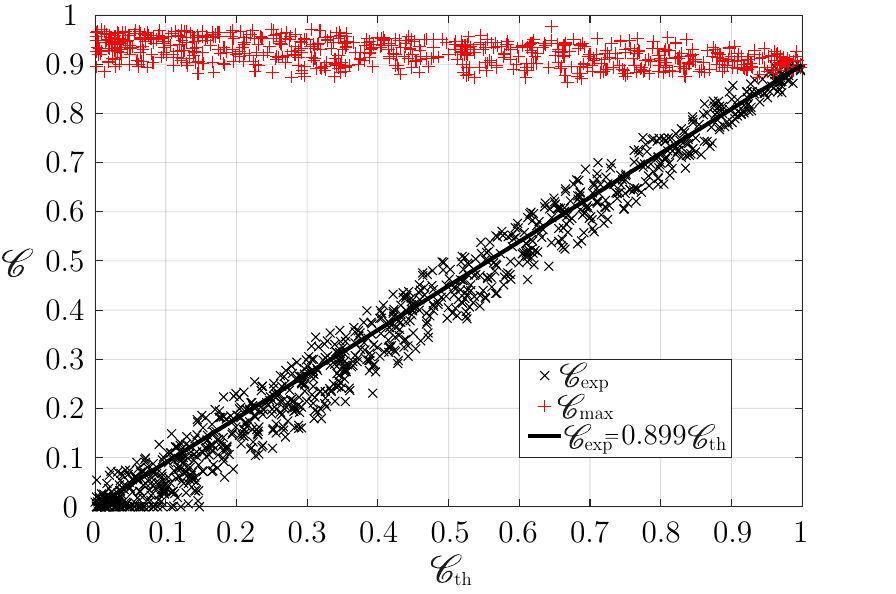}
  \caption{Concurrence $\mathscr{C}_{\exp}$ for a given theoretical $\mathscr{C}_{\rm{th}}$, linear scaling used between $\mathscr{C}_{\exp}$ and $\mathscr{C}_{\rm{th}}$ is also shown as a straight line. $\mathscr{C}_{\max}$ is computed for all positive definite density matrices.}
  \label{fig:C_Cmax}
\end{figure}

In fact, it is possible to quantify the drop of concurrence which is due to the mixedness of the two-qubit state. Indeed, for all mixed states $\rho$ with given purity (i.e. characterized by a given set of eigenvalues $\lambda_1\geq \lambda_2\geq \lambda_3\geq \lambda_4\geq 0$) there is a rigorous upper bound on possible concurrence\,\cite{PhysRevA.64.012316,wooters:eofandconcurrence}
\be
\mathscr{C}_{\max}(\rho) = \max(0,\lambda_1-\lambda_3-2\sqrt{\lambda_2\lambda_4}),
\label{eq:C_max}
\ee
For all of our random states which have positive measured density matrix it is possible to compute $\mathscr{C}_{\max}$. Fig.\,\ref{fig:C_Cmax} displays $\mathscr{C}_{\exp}$ and $\mathscr{C}_{\max}$ as a function of the theoretical concurrence of each generated states $\mathscr{C}_{\rm{th}}$. First we see that $\mathscr{C}_{\exp}$ is fairly linear as a function of $\mathscr{C}_{\rm{th}}$, and that the slope is less than unity as expected. Second, and more interesting, we see that states generated on the high end of the concurrence do reach $\mathscr{C}_{\max}$, showing that achievable purity is the limiting factor for concurrence, and as a result the principal cause of the flattening of the sphere along the $\mathscr{C}$ axis as observed in the raw data (Fig.\,\ref{fig:raw_data_random}). The linear approximation used in the previous section  $\frac{\mathscr{C}_{\exp}}{\mathscr{C}_{\rm{th}}} = 0.899$ is also displayed in Fig.\,\ref{fig:C_Cmax}, and corresponds to the straight line joining the two end points since they are the most relevant (and for enhanced accuracy the highest end point at maximal concurrence ($\alpha=\pi/2,\theta=\pi$) has been computed using a hundred density matrix evaluations).

Concurrence is an entanglement monotone, so one could say that the linearity indicates that the efficiency in the preparation of an entangled state is fairly constant, but limited on the high end by achievable purity. 

\section{Conclusion}
Our work is, to the best of our knowledge, the first experimental explicit verification of the beautiful Jakob-Bergou ``triality'' relation for bipartite pure quantum states of two qubits\,\cite{jakob2003,ref:bergou2010,PhysRevA.76.052107}. This relation really represents the full quantitative completion of local Bohr's complementarity principle by quantum entanglement (concurrence) for this case. The measurements on the two superconducting qubits of the IBM Q quantum computer have shown that the duality of each qubit can indeed be turned off completely, or set to any desired amount by controlling the degree of entanglement between the qubits. Clearly, the Jakob-Bergou relation can be separated into mutually exclusive local and nonlocal parts as
\be
\mathscr{S}_k^2+\mathscr{C}^2=1
\label{eq:C2Sk2 = 1}
\ee
where $\mathscr{S}_k^2=\mathscr{P}_k^2+\mathscr{V}_k^2$ is the amount of locality since the predictability $\mathscr{P}_k$ and the visibility (coherence) $\mathscr{V}_k$ are local with respect to subsystem $k$. Maximal entanglement of the bipartite system ($\mathscr{C}=1$) implies that the local realities must totally disappear ($\mathscr{S}_k=0$), synonym of maximal amount of nonclassical nonlocal phenomena such as violations of Bell inequalities. 

\

Finally, such experiments with the superconducting qubits of the IBM Q quantum computer could be extended in different fundamental directions. First it would be interesting to test a generalization of the Jakob-Bergou relation derived for non-orthogonal alternatives using POVM's in a similar interference and which-state information experiments for two qubits\,\cite{Fonseca2014}:
\be
V^2+P^2+U^2+\mathscr{C}^2 = 1  \, .
\label{eq:fonseca_relation}
\ee
Here $\mathscr{C}$ remains the only purely bipartite quantity as before, but $V,P$ would become the non-orthogonal counterparts of visibility and predictability, and $U$ would be a new single qubit quantity involving the overlap of non-orthogonal markers. Second, the test of generalized triality relation to higher-dimensional systems like qudits (c.f.\,\cite{PhysRevA.76.052107}, \,\cite{Wu2020} and references therein), or even more interestingly to continuous variable systems, would also be one of the next steps (note that qudits would be emulated by collections of qubits on IBMQ). Third, we are also interested in the dynamical evolution of these relations under decoherence, which is inevitable in a quantum computer. For the resulting mixed states it is well-known that Eqs.\,(\ref{eq:jakob_bergou_relation}) and (\ref{eq:fonseca_relation}) become undersaturated inequalities (LHS\,$<1$), but more interestingly the evolution of $\mathscr{C}$ can be surprising, leading for example to entanglement sudden death\,\cite{Almeida2007}, and one may wonder about the comparative evolution of each term. Fourth, one expects that multipartite quantum states which are presently realized on IBM Q quantum computers\,\cite{doi:10.1002/qute.201970031}, and which are essential for applications in quantum information, also possess rich entanglement-separability duality relations of their own, which are of fundamental interest.

\section{Acknowledgements}
\begin{acknowledgements}
We thank Dr. James R. Wootton for useful discussions and comments. We acknowledge use of the IBM Quantum Experience for this work. The views expressed are those of the authors and do not reflect the official policy or position of IBM or the IBM Quantum Experience team.
\end{acknowledgements}
\section{Data availability statement}
The code that supports this study is openly available in \texttt{GitHub} at \url{https://github.com/NicoSchwaller/Duality-and-Entanglement-of-two-Qubits}.
\bibliography{bibliography}
\bibliographystyle{ieeetr}
\appendix
\section{Appendix A: state preparation}
\label{ap:circuit}

The left-hand side of the circuit in Fig.\,\ref{fig:quantum_circuit} prepares the state
$
\ket{\psi}=\cos\frac{\alpha}{2}\ket{00}+\cos\frac{\theta}{2}\sin\frac{\alpha}{2}\ket{10}+\sin\frac{\theta}{2}\sin\frac{\alpha}{2}\ket{11},
$
by applying the unitary transformation
\begin{align*}
&(\rm{CR_y}(\theta)_{A\rightarrow B})(\rm{R_y}(\alpha)\otimes\mathbb{I})=\\
&\begin{pmatrix} 
1 & 0 & 0 & 0  \\
0 & 1 & 0 & 0  \\
0 & 0 & \cos\frac{\theta}{2} & -\sin\frac{\theta}{2}  \\
0 & 0 & \sin\frac{\theta}{2} & \cos\frac{\theta}{2}  \\
\end{pmatrix}
\left[\begin{pmatrix} 
\cos\frac{\alpha}{2} & -\sin\frac{\alpha}{2}  \\
\sin\frac{\alpha}{2} & \cos\frac{\alpha}{2}  \\
\end{pmatrix}
\otimes
\begin{pmatrix} 
1 & 0  \\
0 & 1  \\
\end{pmatrix}\right].
\end{align*}

\begin{figure}[H]
\centering
  \includegraphics[width=1\linewidth]{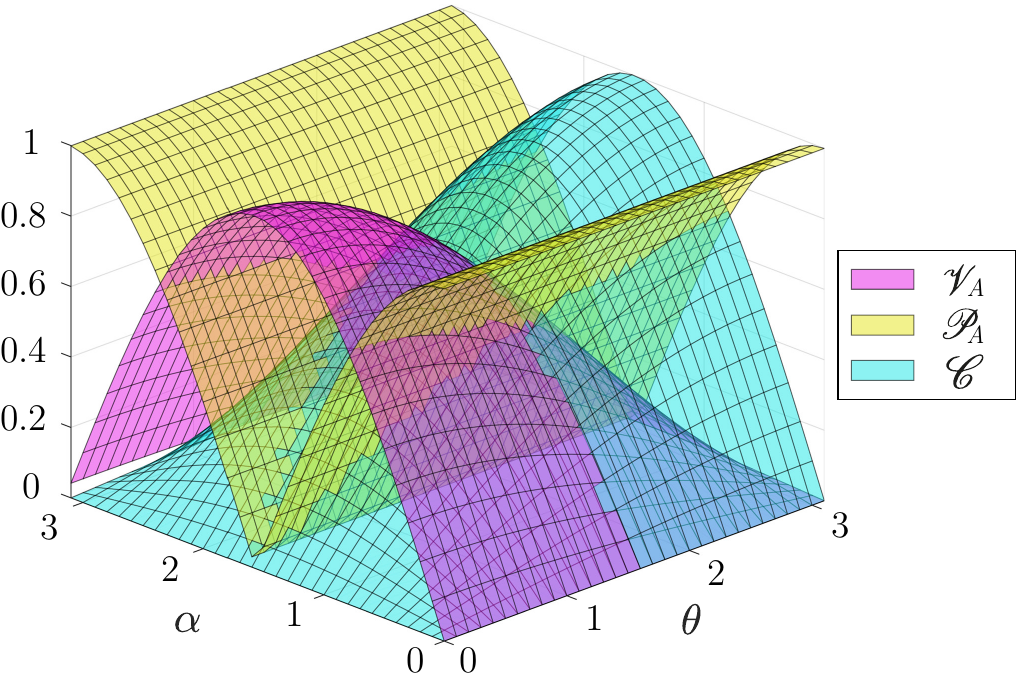}
  \caption{Analytical values of $\mathscr{V}_A$, $\mathscr{P}_A$ and $\mathscr{C}$ as a function of $\alpha$ and $\theta$.}
  \label{fig:ExpscanVPC}
\end{figure}

According to equations (\ref{eq:def_C}\,-\,\ref{eq:def_P}), such a unitary operation acting on the state $\ket{00}$ allows the five quantities $\mathscr{V}_k$, $\mathscr{P}_k$ (with $k=A,B$) and $\mathscr{C}$ to reach their extremal values, i.e. 0 and 1, as shown in Fig.\,\ref{fig:ExpscanVPC} for $k=A$.
\begin{table}[H]
\centering
\begin{tabular}{c|c|c}
\textbf{State} & $\alpha$     & $\theta$ \\ \hline
\textbf{1}     & $\pi/2$       & $\pi$    \\
\textbf{2}     & 1.0472       & $\pi$    \\
\textbf{3}     & 1.1230       & 3.6216   \\
\textbf{4}     & \,\,\, 1.3181 \,\,\,       & \,\,\, 4.0016 \,\,\,  \\
\textbf{5}     & $\pi/2$       & 4.0816   \\
\textbf{6}     & 0.5236       & $\pi$    \\
\textbf{7}     & 0.7247       & 4.4416   \\
\textbf{8}     & 1.1230       & 5.1196   \\
\textbf{9}     & $\pi/2$       & 5.1050   \\
\textbf{10}    & 0            & 0        \\
\textbf{11}    & 0.5236       & 0        \\
\textbf{12}    & 1.0472       & 0        \\
\textbf{13}    & $\pi/2$       & 0       
\end{tabular}
\caption{Couples of values $(\alpha,\theta)$ used to prepare the states of Fig.\,\ref{fig:VDCsim} with our circuit.}
\label{tab:values_alpha_theta}
\end{table}
\section{Appendix B: raw values in second experiment}
\label{ap:data}
\begin{figure}[H]
\multirow{2}{*}{}{
         \begin{subfigure}[h]{0.5\textwidth}
             \includegraphics[width=1\linewidth]{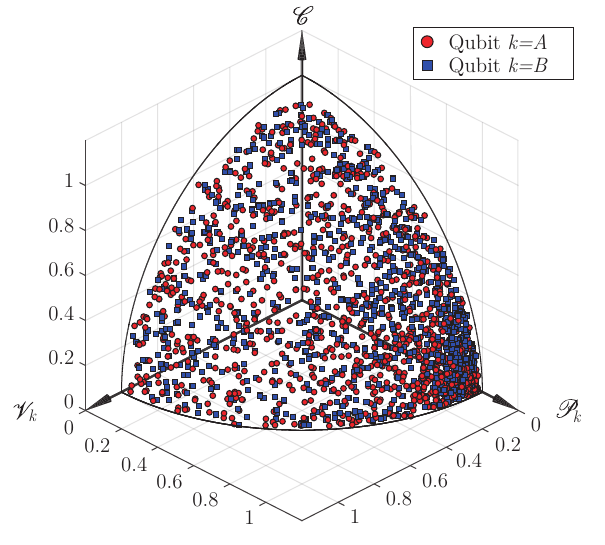}
             \caption{\centering}
             \label{fig:raw_data_random}
             
         \end{subfigure}\\
         \begin{subfigure}[h]{0.5\textwidth}
             
             \includegraphics[width=1\textwidth]{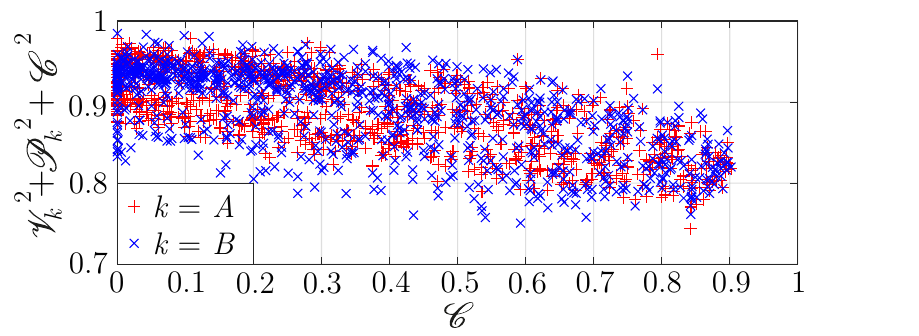}
             \caption{\centering}
             \label{fig:raw_8192_triality}
         \end{subfigure}\\
         \begin{subfigure}[h]{0.5\textwidth}
             
             \includegraphics[width=1\textwidth]{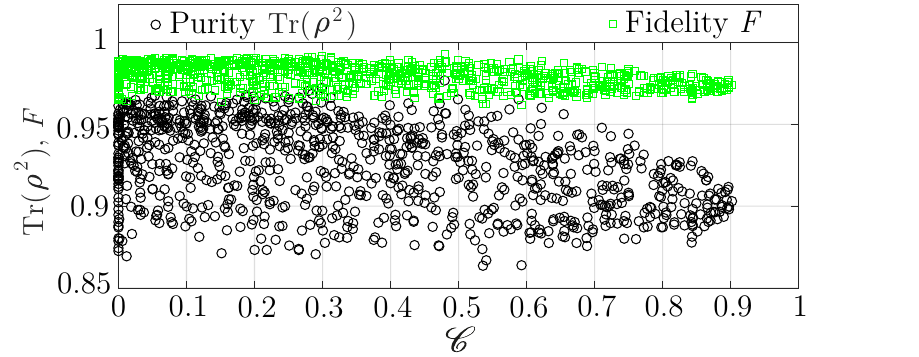}
             \caption{\centering}
             \label{fig:raw_8192_PF}
         \end{subfigure}\\
         }
\caption{Raw values for 1000 states generated with random ($\alpha,\theta$) values\,\cite{date_meas_8192} of (a) the triality relation for qubits $k=A,B$ (b) the triality relation for $k=A$, as a function of $\mathscr{C}$ and (c) the purity and fidelity of the two-qubit states.}
\label{fig:Raw_data}
\end{figure}

Fig.\,\ref{fig:raw_data_random} reports the raw values of $\mathscr{V}_k$ and $\mathscr{P}_k$ ($k=A, B$), as well as $\mathscr{C}$, after 8192 shots per circuit, for 1000 two-qubit states generated with the circuit of Fig.\,\ref{fig:quantum_circuit} using random values of $\alpha, \theta$ uniformly sampled in $[0,\pi]$ (implying an irrelevant slight oversampling in the right corner of the octant). All points are clearly interior to the unit sphere, showing that the Jakob-Bergou triality inequality (\ref{eq:VPC_leq_1}) is satisfied everywhere for both qubits $A,B$, and that 8192 shots are sufficient to eliminate all violations.
Fig.\,\ref{fig:raw_8192_triality} displays more in detail the triality relation for both qubits $A,B$ as a function of concurrence. 

On Fig.\,\ref{fig:raw_8192_PF} we plot the purity $\Tr(\rho_{\exp}^2)$ and the fidelity   
\be
F=\Tr\left(\sqrt{\sqrt{\rho_{\rm{th}}}\rho_{\exp}\sqrt{\rho_{\rm{th}}}}\right)^2
\label{eq:Fidelity}
\ee
of each of the measured thousand states $\rho_{\exp}$ respectively to its target $\rho_{\rm{th}}$. Not astonishingly purity and fidelity decrease with concurrence, the latter being limited by purity (c.f. Fig.\,\ref{fig:C_Cmax} and its discussion in the main text).

\end{document}